# An atomic-resolution nanomechanical mass sensor

K. Jensen, Kwanpyo Kim & A. Zettl

*Department of Physics, University of California at Berkeley,*

*Center of Integrated Nanomechanical Systems, University of California at Berkeley,*

*Materials Sciences Division, Lawrence Berkeley National Laboratory,*

*Berkeley, CA 94720, U.S.A.*

**Mechanical resonators are widely used as inertial balances to detect small quantities of adsorbed mass through shifts in oscillation frequency[1]. Advances in lithography and materials synthesis have enabled the fabrication of nanoscale mechanical resonators[2-6], which have been operated as precision force[7], position[8,9], and mass sensors[10-15]. Here we demonstrate a room temperature, carbon nanotube-based nanomechanical resonator with atomic mass resolution. This device is essentially a mass spectrometer with a mass sensitivity of $1.3 \times 10^{-25}$ kg/√Hz, or equivalently, 0.40 gold atoms/√Hz. Using this extreme mass sensitivity, we observe atomic mass shot noise, which is analogous to the electronic shot noise[16,17] measured in many semiconductor experiments. Unlike traditional mass spectrometers, nanomechanical mass spectrometers do not require the potentially destructive ionization of the test sample, are more sensitive to large molecules, and could eventually be incorporated on a chip.**

Nanomechanical resonators function as precision mass sensors because their resonant frequency, which is related to their mass, shifts when a particle adsorbs to the resonator and significantly increases the resonator's effective mass. In general, the relation between shifts in resonant frequency and changes in mass depends on the geometry of the resonator and the location of the adsorbed particle. For a cantilevered beam resonator, this relation is described by a responsivity function, $R(x)$, which is defined as the ratio of the shift in resonant frequency, $\Delta f$, to the change in mass, $\Delta m$, as a function of position, $x$, of the adsorbed mass along the beam. (An approximate form for $R(x)$ is derived in the Supplementary Information.) Assuming that the adsorbed mass is distributed evenly along the resonator, this relation can be simplified by averaging over the responsivity function to obtain

$$\Delta f = \overline{R(x)} \cdot \Delta m = -\frac{f_0}{2m_0} \Delta m \tag{1}$$

where $f_0$ is the resonant frequency of the beam and $m_0$ is the initial mass of the beam.

In order to maximize the magnitude of the responsivity, it is apparent from equation (1) that reducing the mass of the resonator, while maintaining high resonance frequencies, is critical. Carbon nanotubes are ideally suited for this task. They are naturally much smaller and less dense than resonators manufactured using standard e-beam lithographic techniques, and thus their mass (~$10^{-21}$ kg) is typically more than four orders-of-magnitude less than state-of-the-art micromachined resonators (~$10^{-17}$ kg) [13]. Finally, because of their high elastic modulus[18], even small, slender nanotubes maintain high resonance frequencies. As a result of such properties, carbon nanotubes have previously been used to detect changes in mass as small as $10^{-21}$ kg[14].

Another consideration is the geometry of the nanomechanical resonator. Although many previous attempts at precision mass sensing have focused on doubly clamped geometries[10,13-15] to allow simple electrical readout, singly clamped geometries have notable advantages. Their dynamic range, essentially how far they can bend before non-linear effects dominate, is significantly increased. Also, singly clamped resonators tend to have higher quality factors (i.e. sharper resonance peaks) due to reduced clamping losses[19]. Our quality factors were typically on the order of 1000. Both dynamic range and quality factor are important in determining a resonator's ultimate sensitivity[11].

A transmission electron microscope (TEM) image of a typical nanotube-based nanomechanical mass spectrometer device is shown in Fig. 1a. The entire device consists of a single arc-grown double-walled nanotube[20] attached to one electrode in close proximity to a counter electrode (not shown). We chose double-walled nanotubes over smaller single-walled nanotubes because of their increased rigidity and uniform electrical properties (i.e. mostly metallic). Fabrication of these devices is described in detail in previous work[21].

These high resolution TEM images enable precision calibration of our devices through determination of their exact size and thus mass. A double-walled nanotube's mass is simply $m_{CNT}=2 \cdot m_C \cdot \pi \cdot (D_i+D_o) \cdot L/A_{gr}$ where $m_C$ is the mass of a carbon atom, $D_i$ and $D_o$ are the inner and outer shell diameters, $L$ is the length, and $A_{gr}$ is the area of graphene's unit cell. For the device shown in Fig. 1a ($D_i$=1.75 nm, $D_o$=2.09 nm, $L$=254 nm), $m_{CNT}=2.33\times10^{-21}$ kg. After calibration by TEM, the nanotube device is transferred to an external measurement apparatus.

The physical layout of the entire nanomechanical mass spectrometer apparatus, including nanotube device and evaporation system, is shown in Fig. 1b. The nanotube device is placed at one end of an ultra-high vacuum (UHV) chamber ($10^{-10}$ torr). To load atoms onto the device, we evaporate gold from a tungsten filament a distance $d_{CNT}$=50.2 cm away from the nanotube device. A shutter may be inserted between the evaporation source and the nanotube to interrupt the gold mass loading. A water-cooled quartz crystal microbalance (QCM), a distance $d_{QCM}$=12.8 cm from the evaporation source and normal to the direction of evaporation, is used as a secondary means of calibrating the nanotube device.

One difficulty of using nanomechanical resonators as precision sensors is the detection of the mechanical vibrations of the resonator. We use a detection technique based on a nanotube radio receiver design[21]. In effect, we broadcast a radio signal to the nanotube and listen for its vibrations. This technique relies on the unique field emission properties of carbon nanotubes[22], one of which is a strong coupling between the field emission current and the nanotube's mechanical vibrations. A schematic for the electrical detection circuit is shown in Fig. 1c.

In a typical experiment, we adjust the gold evaporation source's filament current, with the shutter closed, until we measure a steady mass flux on the QCM. We then open and close the shutter multiple times, loading a small number of gold atoms onto the nanotube each time. As expected, the resonant frequency of the nanotube shifts downward during evaporation and remains steady with the shutter closed. The resonant frequency of the nanotube is automatically tracked and recorded at a sampling rate typically between 10 and 100 Hz.

Data from such an experiment are shown in Fig. 2. Here white regions indicate that the shutter is open, while shaded regions indicate that the shutter is closed, blocking the gold atoms. The nanotube used in this particular experiment has geometry and mass, determined from TEM images, described by the following parameters: $D_o$=1.78 nm, $D_i$=1.44 nm, $L$=205 nm, $m_{CNT}$=1.58×10$^{-21}$ kg. The initial resonant frequency of the nanotube was set near $f_0$=328.5 MHz through electrostatic tensioning[21,23]. From the resonant frequency and mass of the resonator, we expect a responsivity of 0.104 MHz/zg (1 zg = 10$^{-24}$ kg). A scale converting frequency shift to mass using this responsivity is shown on the vertical axis to the right. According to this scale, the frequency shift in the first "open" section corresponds to just 51 gold atoms.

We now examine how to use our sensitive mass sensor as a mass spectrometer. The noise on the plateaus in Fig. 2a, when no atoms are loaded on the nanotube, demonstrates that the sensitivity of our device is 0.13 zg/√Hz or equivalently 0.40 Au atoms/√Hz. This is the lowest mass noise ever recorded for a nanomechanical resonator, which is even more striking considering that this measurement was performed at room temperature rather than in a cryogenic environment. These noise levels clearly indicate that we have achieved atomic sensitivity. However, to determine the mass of an adsorbed atom, it is also necessary to know, along with the resulting frequency shift, the position of the atom along the nanotube. One method of accomplishing this is to occlude portions of the resonator so that atoms must land at a specific location. Another method, which we employ here, relies on the statistics of the frequency shifts.

Atoms arrive at the nanotube at a constant average rate. However, because atoms are discrete, the number arriving during any given time interval is governed by Poisson statistics. This effect can be seen in Fig. 2a where the adsorption rates for the four open sections show significant variation (2.2, 2.1, 3.5, and 2.6 Au atoms/s). There are two independent approaches of using Poisson statistics to measure the mass of the gold atoms. The first approach relies on measuring statistical fluctuations in mass adsorption rate, which we term atomic mass shot noise. The second approach analyzes the statistical distribution of frequency shifts that occur each sampling period. We now consider each approach in turn.

In analogy with electronic shot noise, which has spectral density $S_I^{(shot)}(f) = 2eI$ [24], we expect that the mass adsorption rate will have atomic mass shot noise with spectral density $S^{(shot)}_{\frac{dm}{dt}}(f) = 2m_{Au}\frac{dm}{dt}$. Here we are making a few reasonable assumptions, which are supported by previous studies. We assume that gold arrives as single atoms rather than in clusters[25]. We assume that the arrival of gold atoms is uncorrelated. Finally, we assume that, after landing on the nanotube, gold atoms find a nucleation site to adhere to in a time short compared to the measurement time[26].

An additional complication arises from the fact that our experiment does not measure mass adsorption rate directly, but rather measures the time derivative of the resonant frequency, which is related to the mass adsorption rate though the responsivity function of the resonator, $R(x)$. To account for $R(x)$, we sum the noise contribution at each point along the resonator and arrive at the final equation for atomic mass shot noise:

$$S^{(shot)}_{\frac{df_0}{dt}}(f) = 2m_{Au}\frac{dm}{dt} \cdot \frac{1}{L}\int_0^L dx[R(x)]^2 \approx 1.17\frac{f_0^2}{m_{CNT}^2}m_{Au}\frac{dm}{dt}. \quad (2)$$

Besides atomic mass shot noise, there are other significant noise sources such as readout noise and thermal-mechanical noise[11]. Both of these noise sources are frequency independent, or white, in $S_{f_0}(f)$. Thus, they will appear as *differentiated* white noise, which grows as the square of the measurement frequency[27], in $S_{\frac{df_0}{dt}}(f)$.

Figure 3a shows noise levels in our measurements as a function of measurement frequency. At higher frequencies, some form of differentiated white noise, such as readout noise, dominates. However, at lower frequencies for the evaporation case, atomic mass shot noise dominates. The parameters from our experiment yield an expected atomic mass shot noise of 0.016 MHz$^2$/s$^2$/Hz, which is drawn as the horizontal gray line in the figure. The total expected noise, including the measured differentiated white noise, is drawn as the dark black line. The data for the evaporation case follow the expected noise level well. A fit to the data yields a measured atomic mass shot noise of 0.014±0.002 MHz$^2$/s$^2$/Hz, which would result from an atomic mass of 0.29±0.05 zg, consistent with the accepted mass of gold, 0.327 zg. Thus, we have successfully determined the mass of gold atoms with a nanomechanical resonator.

The low frequency noise for the shuttered case deviates somewhat from differentiated white noise indicating that another low frequency noise process exists, which does not depend on the evaporation of atoms. However, in this case the noise is an order-of-magnitude less than atomic mass shot noise, indicating that we have sufficient long term stability for our shot noise measurements. A potential explanation for this noise source is the current-induced motion of atoms along the surface of the resonator, which may be controlled by limiting the current to sufficiently low levels [28].

We now turn to the statistical distribution of frequency shifts that occur each sampling period. The histogram of frequency shifts, shown in Fig. 3b, provides additional evidence of the Poissonian nature of the mass adsorption process and also provides an independent, but related, means of determining the mass of atoms. Assuming a constant evaporation rate, an approximately uniform distribution of atoms along the resonator, and Gaussian noise sources, it is possible to derive the expected distribution of frequency shifts, shown as the black line in Fig. 3b (see Supplementary Information for details.) Due to Poisson statistics, the shape of the expected distribution depends on the number of atoms that adsorb to the resonator per sampling period. Because the mass adsorption rate is well known, the number of atoms per sample depends on the mass of a single atom. The inset shows a measure, based on a $\chi^2$ test, of how well the data fit the expected distribution calculated for various values of atomic mass. Due to the number of large downward frequency shifts ($\Delta f$<-100 kHz), these data are only consistent with distributions calculated for an atomic mass between 0.1 zg and 1 zg ($m_{Au}$=0.327 zg). This is an independent measurement of the mass of a gold atom, though the atomic mass shot noise technique is more precise. Of course the underlying purpose of this work is not to obtain a revised value for the mass of a gold atom, but rather to demonstrate the power of the technique.

Our nanomechanical mass spectrometer has significant advantages over traditional high-resolution mass spectrometers. Most notably, it does not require ionization of the test sample, which makes it more suitable for large biomolecules such as proteins. These molecules are often destroyed during ionization even with "soft" ionization techniques such as matrix-assisted laser desorption/ionization (MALDI )[29] and electrospray ionization (ESI)[30]. Our device becomes more sensitive at higher mass ranges, in contrast with traditional mass spectrometers. Finally, our device is compact, as it does not require large magnets or long drift tubes, and could in principle be incorporated on a chip.

**Methods**

The QCM provides an alternate method of calibrating the responsivity of the nanotube mass spectrometer, which was initially calculated from TEM-determined parameters. Of course, the QCM does not have the sensitivity to weigh single atoms; however because it averages over a relatively large area, it is an excellent means of measuring mass flux. The mass adsorption rate at the nanotube calculated from the mass flux at the QCM assuming an isotropic evaporation source is

$$\frac{dm_{CNT}}{dt} = \alpha \cos\theta_{CNT} \frac{d_{QCM}^2}{d_{CNT}^2} A_{CNT} \frac{1}{A_{QCM}} \frac{dm_{QCM}}{dt}. \qquad (3)$$

Only the sticking coefficient of gold on a nanotube, $\alpha$, and the misalignment angle of the nanotube to the evaporation source, $\theta_{CNT}$, are not precisely known. Fortunately, it is simple to extract these parameters by varying the evaporation rate. The inset in Fig. 2a shows the rate of frequency change of the nanotube resonator as a function of mass flux at the QCM over multiple experimental runs. A fit to the data gives a ratio between these quantities of 2.18±0.13 MHz nm$^2$/zg. This implies that $\alpha \cos\theta_{CNT} = 0.88\pm0.06$, which is reasonable assuming a well aligned nanotube and a relatively high sticking coefficient[26].

Using equation (3) and the experimentally determined value of $\alpha \cos\theta_{CNT}$ it is possible to calculate the mass adsorption rate at the nanotube. The QCM records a constant evaporation rate of 2.44 ng/s, which corresponds to an adsorption rate of 1.01 zg/s or equivalently 3.09 Au atoms/s at the nanotube. In comparison, the adsorption rate, calculated using the TEM determined responsivity, is 2.94 Au atoms/s. After accounting for uncertainties in our measurement of $\alpha \cos\theta_{CNT}$ and for natural, Poissonian variations in adsorption rate, these values are in agreement, and thus, the measurements from the QCM are consistent with the TEM determined responsivity.

**Author Contributions** K.J. and A.Z. conceived the experiments and co-wrote the paper. K.J. designed and constructed the experimental apparatus, prepared nanotube samples, recorded the data, and analyzed the results. K.K. aided with apparatus construction and sample preparation.

**Acknowledgements** We thank B. Aleman for technical assistance. This work was supported by the Director, Office of Energy Research, Office of Basic Energy Sciences, Materials Sciences and Engineering Division, of the U.S. Department of Energy under Contract No. DE-AC02-05CH11231, which provided for nanotube synthesis, detailed TEM characterization, and UHV testing of the nanomechanical mass spectrometer, and by the National Science Foundation within the Center of Integrated Nanomechanical Systems, under Grant No. EEC-0425914, which provided for design and assembly of the spectrometer. K.K acknowledges support from a Samsung Graduate Fellowship.

**Figure 1 | Nanomechanical mass spectrometer device and schematics. a**, TEM images of a nanomechanical mass spectrometer device constructed from a double-walled carbon nanotube. From these high resolution TEM images, the geometry and thus mass of the nanotube are precisely determined ($m_{CNT}$=2.33×10$^{-21}$ kg), which is an essential calibration for the mass spectrometer. **b,** Physical layout of the entire nanomechanical mass spectrometer apparatus. Gold atoms are evaporated, inside a UHV chamber, and travel a distance $d_{CNT}$ before adsorbing to the nanotube device and consequently lowering its resonant frequency. A shutter may be inserted to interrupt mass loading. The QCM provides an alternate means of calibrating the system through measurement of mass flux. **c**, Schematic of the mechanical resonance detection circuit. Briefly, the electrode opposite the nanotube is biased to induce a field emission current from the nanotube. An

amplitude modulated (AM), frequency-swept, via a voltage-controlled oscillator (VCO), RF signal is coupled to the nanotube forcing it into resonance, and consequently modulating the field emission current. The modulated field emission current is recovered by a lock-in amplifier and the resonance peak is displayed on the oscilloscope or recorded by a computer.

**Figure 2 | Frequency shifts during mass loading. a**, The nanotube's resonant frequency (left y-axis) and change in adsorbed mass (right y-axis) versus time during evaporation of gold. The resonant frequency shifts downward when the shutter is open (white regions) and remains constant when the shutter is closed and blocking the gold atoms (shaded regions). The frequency shift in the first open section corresponds to just 51 gold atoms adsorbing to the nanotube. The inset shows a calibration, discussed in the Methods, of the nanotube's frequency shift rate versus the mass flux at the QCM. **b**, At the same time, the QCM records a constant evaporation rate as demonstrated by the constant slope of the frequency shift (left y-axis) and change in mass (right y-axis). Notably, the mass deposited on the QCM is measured in nanograms versus the zeptogram scale used for the nanotube.

**Figure 3 | Atomic mass shot noise. a**, Spectral density of the noise in the time derivative of the resonant frequency during evaporation (red) and when the shutter is closed (blue). Shaded regions indicate uncertainty in the estimation of spectral density. In both cases, differentiated white noise dominates at higher frequencies. At lower frequencies for the evaporation case, there is a significant increase in noise caused by the discrete nature of the arrival of mass

(i.e. atomic mass shot noise). The horizontal gray line depicts the predicted level of atomic mass shot noise, and the sloped gray line is the measured value of differentiated white noise. The black line is the sum of these noise sources. **b**, Histogram of frequency shifts per sampling time during evaporation. The black line shows the expected number of counts according to our model given the correct value of the atomic mass, $m_{Au}$. The inset shows a measure of how well the data fit the expected distribution calculated for various values of atomic mass. Due to the large number of downward frequency shifts ($\Delta f$<-100 kHz), the data are only consistent with distributions calculated for atomic masses between 0.1 zg and 1 zg ($m_{Au}$ = 0.327 zg).

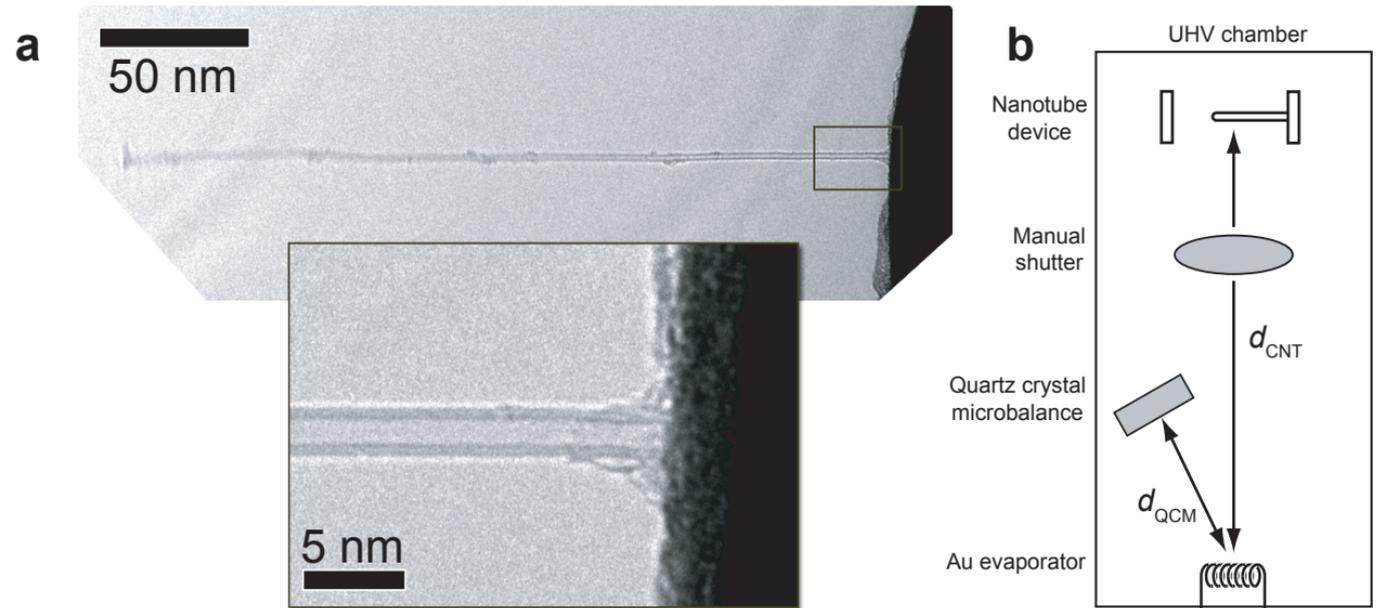
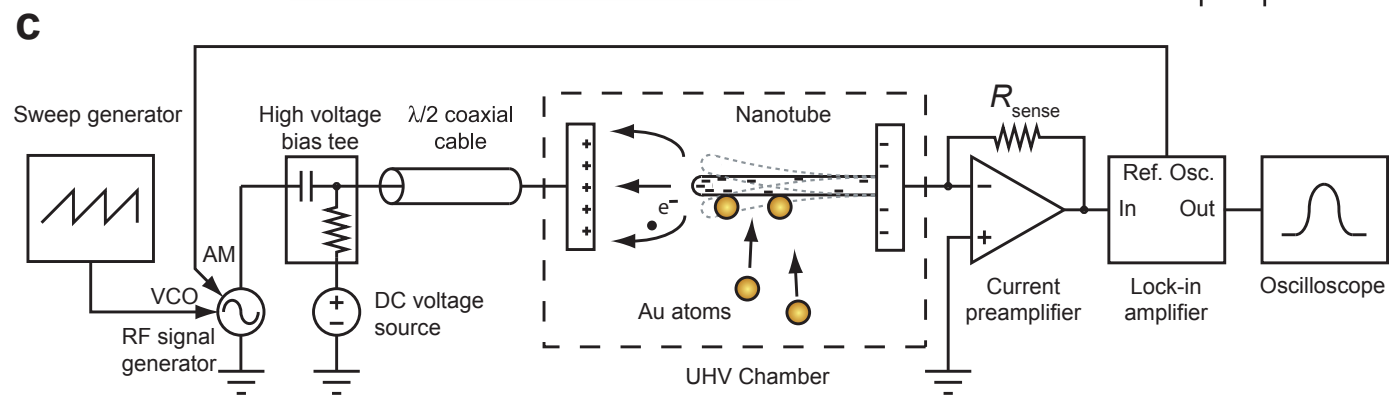

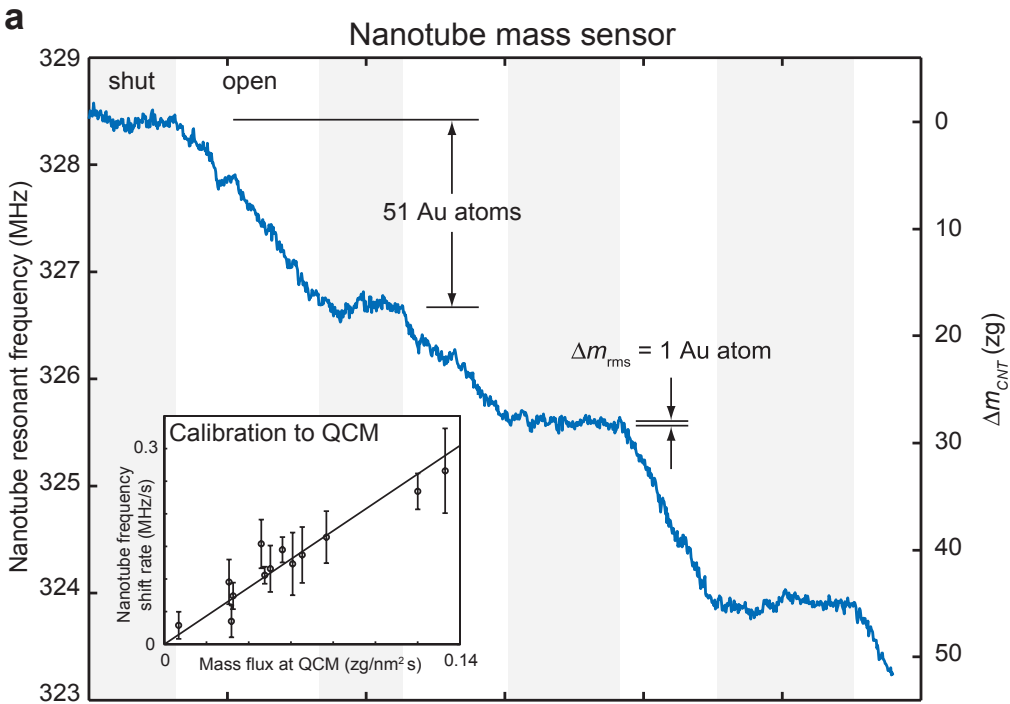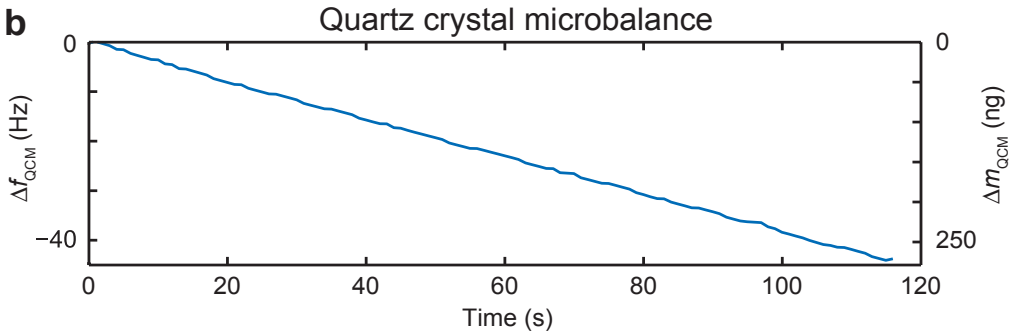

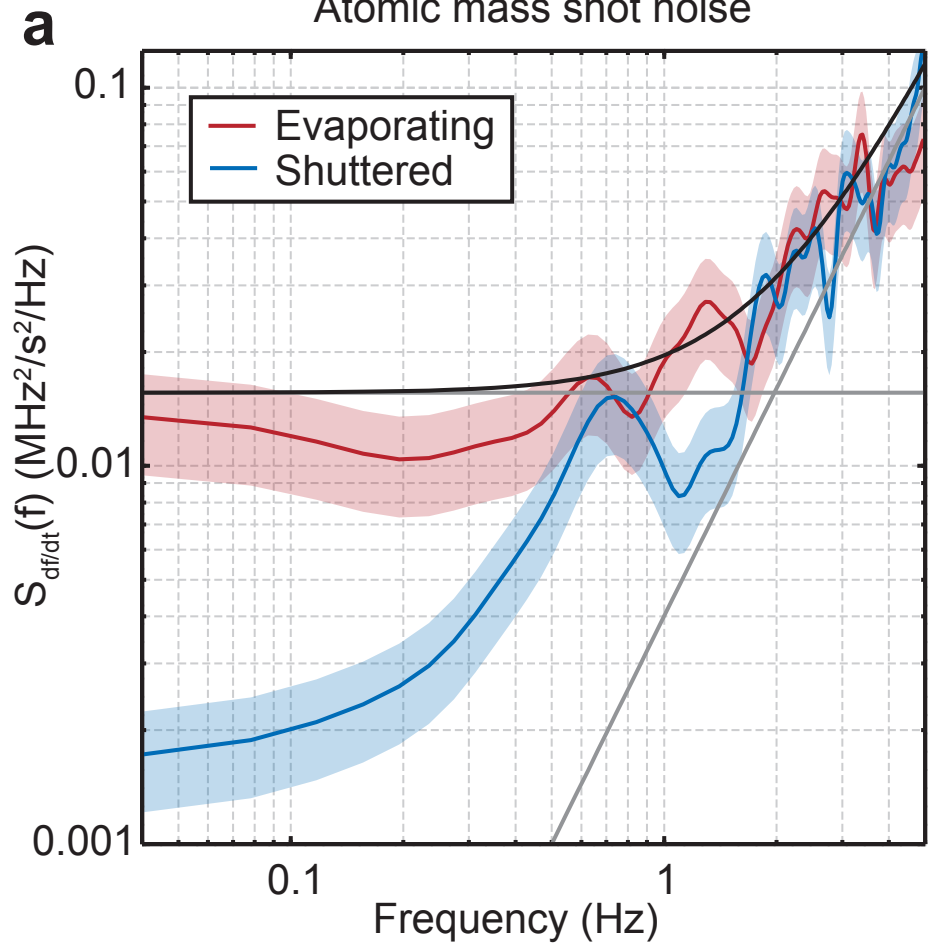 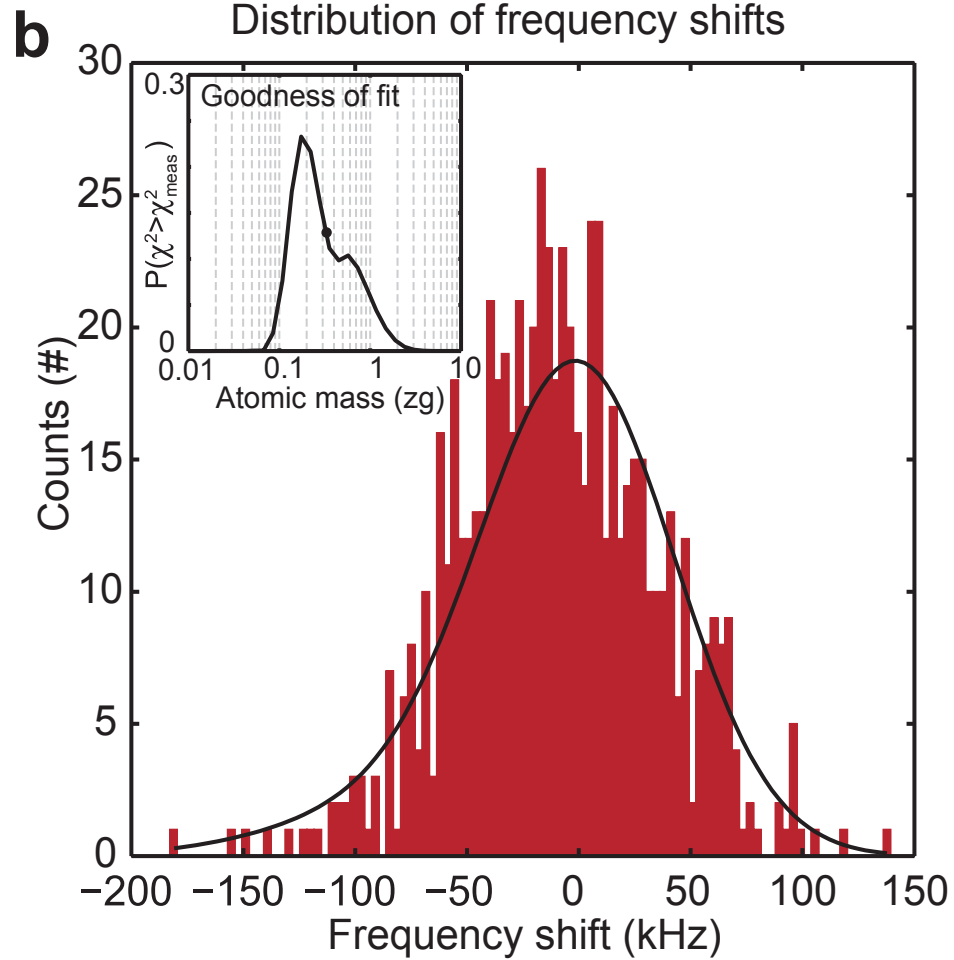